\documentstyle[multicol,psfig,epsf,aps,pre,eqsecnum]{revtex}
\begin{document}
\draft
\tightenlines
\title{\bf{ Reunion of random walkers with a long range interaction:
          applications to  polymers and quantum mechanics}}
\author{Sutapa Mukherji\cite{eml1}} 
\address{ Department of Physics and Meteorology, Indian Institute of
  Technology, Kharagpur 721 302, India}
\author{ Somendra M. Bhattacharjee\cite{eml2}} 
\address{Institute of Physics, Bhubaneswar 751 005, India\\
Dipartimento di Fisica, Universit\`a di Padova, Via 
Marzolo 8,  35131 Padova, Italy}
\onecolumn
\date{\today}
\maketitle
\widetext
\begin{abstract}
We use renormalization group to calculate the reunion and survival
exponents of a set of random walkers interacting with a long range
$1/r^2$ and a short range interaction.  These exponents are used to
study the binding-unbinding transition of  polymers and the
behavior of several quantum problems. 
\end{abstract}
\begin{multicols}{2}
\section{Introduction}
\label{sec:Introduction}
The problem of reunion of interacting random walkers appears in many
situations, directly or in disguise, such as e.g., in situations involving
one-dimensional stringlike objects such as interfaces, steps on
surfaces, in one-dimensional quantum problems, in several types of
polymer problems, wetting, etc. The issue quite often is the
large-length scaling of the partition function of reunion (or
survival) of a set of walkers all starting, say, at origin.  For a
purely short range repulsive interaction, this is the problem of
``vicious walkers.''  Many of the vicious-walker reunion and survival
exponents are known from exact calculations in one
dimension\cite{fisher84,huse84}, the renormalization-group (RG)
approach\cite{smjpa,smpre} for general dimensions, and lattice models
in two dimensions\cite{gutt}.  Recent studies of phase transitions on
vicinal or miscut surfaces\cite{smb96,smprl99} required the scaling
behavior of such partition functions in the presence of a long range
interaction (namely $1/r^2$) in addition to short range
interactions. The effective interaction of steps on a vicinal surface
has a long $r^{-2}$ tail\cite{kohn} and recent studies on Si(113)
surfaces indicate the presence of an attractive short range
interaction\cite{song94} as well.  Similar long range interactions
occur in the one-dimensional Calogero-Sutherland model\cite{suth} of
interacting quantum particles, and as the angular momentum term in
higher-dimensional quantum problems. {\it The reunion and survival
behaviors turn out to be the unifying feature in these wide varieties
of problems}.  These motivated us to study the scaling limit of various
partition functions in the presence of long range interactions, especially
$r^{-2}$.  We use these results for studying phase transitions of
polymers and then show how many results of quantum problems can be
recovered from such an analysis.

The success of the RG approach for short range
interactions\cite{smjpa,smpre,gutt,smbjj} and the possibility of exact
renormalization in the presence of long range
interactions\cite{kolo,smprl99} allowed us to use the
renormalization-group approach for this problem.  It is known that the
probability 
distribution function for a random walker can be represented by the
Wiener integral, which as a path integral could, {\it a la }Edwards,
represent a polymer.  Any interaction of two or more walkers at the
same time or step length would translate into an interaction among the
polymers at the same contour length.  For $d$-dimensional walkers, we
would then get $(d+1)$-dimensional interacting directed polymers.  We
shall be using this polymer language throughout this paper.

Phase transitions in polymerlike systems by interactions at the same
contour length occur, e.g., in the case of steps already mentioned, in
flux lines in superconductors\cite{sm97}, in melting and unzipping of
DNA\cite{dna1,dna2,dna3,dna4,dna5,dna6}, to cite a few.  A directed
polymer formulation then becomes quite natural.  In this paper, our
focus is mainly on the effect of the long range interaction on these
phase transitions, recovering the results for the short range case as
a special case.    Furthermore, we 
exploit the relation between polymers and quantum
problems to show the importance of reunion behaviors in wide varieties
of problems in various dimensions\cite{sm94}.

In this paper, we recover several results known from exact solutions,
illustrating how the reunion exponent armed with the renormalization group
gives a unified approach to these problems.  The RG approach can be
extended to other types of interactions, especially other marginal
interactions, for which exact solutions are not known, such as that 
required in the vicinal surface problem of Ref. \cite{smprl99}.  There
lies the merit of the approach of this paper.

The Hamiltonian and the general exponents of interest are introduced
in Sec. II.  The general RG approach is discussed in Sec. III,
where the exponents are also calculated, with some of the details
in the Appendix.  These results are then used for polymer
problems in Sec. IV and to quantum problems in Sec. V.  The
conclusion is given in Sec. VI.

\section{Hamiltonian and the exponents}
\label{sec:Hamilt-expon}
The Hamiltonian for $p$-directed polymers with
pairwise interaction\cite{smb96,smprl99} is 
\begin{mathletters}
\begin{equation}
  \label{eq:1}
H_p= \sum_i^p   \int_0^N \! \!\!dz \,  \frac{1}{2} 
\left( \frac{\partial {\bf r}_i(z)}{\partial z}\right )^{^2} +
\sum_{i>j} \int_0^N \!\!\!\! dz \  V{\bf (}{\bf r}_{ij}(z){\bf )},
\end{equation}
where ${\bf r}$ is a $d$ dimensional position vector and ${\bf
r}_{ij}(z) = {\bf r}_i(z) - {\bf r}_j(z)$.  The interaction potential
is taken as
\begin{equation}
  \label{eq:29}
  V({\bf r})  = v_0 \ \delta_{\Lambda}({\bf
  r}) + \frac{g}{ \mid {\bf r}\mid^{\sigma}},
\end{equation}
and $\delta_{\Lambda}({\bf r})$ is a delta function with a cutoff such
that in Fourier space
\begin{eqnarray}
  \label{eq:16}
  \delta_{\Lambda}({\bf q}) &=& 1 \quad {\rm for}\quad q\le \Lambda,
\nonumber\\
&=& 0 \quad {\rm otherwise}.
\end{eqnarray}
\end{mathletters}
We shall use this cutoff for the whole potential.
Our interest is in the asymptotic behavior of the following partition 
functions.\\
(a)  Reunion of all the  chains of length $N$ at a point ${\bf  r}$,
\begin{mathletters}
\begin{eqnarray}
  \label{eq:2}
  &\displaystyle{Z_{{\rm R},p}(N,{\bf r})}& \nonumber \\
&=&\int {\cal D} R \ e^{-H_p} \prod_{i=1}^{p}
  \left [\delta^d {\bf (}{\bf r}_i(0){\bf )}\ \delta^d {\bf (}{\bf
      r}_i(N) - {\bf r}{\bf )} \right ]\\
&\sim& N^{-\psi_{R,p}};
\end{eqnarray}
(b) reunion anywhere,
\begin{eqnarray}
  \label{eq:2b}
   {\cal Z}_{{\rm R},p}(N) &=& \int d^dr \  Z_{R,p}({N,\bf r})\\
&\sim& N^{-\Psi_{R,p}};
\end{eqnarray}
and (c) survival,
\begin{eqnarray}
  \label{eq:2c}
  Z_{{\rm S},p}(N) &=& \int {\cal D} R \  e^{-H_p} \prod_{i=1}^{p}
  [\delta^d ({\bf r}_i(0))]\\
&\sim& N^{-\psi_{S,p}},
\end{eqnarray}
\end{mathletters}
where $\int {\cal D} R$ stands for the summation over all possible
configurations of the polymers (path integrals). The constraint that
{\it all chains are tied together at the origin} is represented by the
product of the $\delta$ functions.  In case (c), the ends at $z=N$ are
free for all the chains.

In the above equations, the asymptotic behaviors define the basic
exponents $(\psi_{R,p},\Psi_{R,p},\psi_{S,p})$ of interest in this paper.
These exponents are expected to be universal, independent of the
detailed microscopic form of the polymers. Hence the choice of a
continuum model.

The notation introduced above is used in Sec. III, where these
exponents are calculated.  However, in the subsequent sections on
polymers and quantum problems, we need mainly $\Psi_{R,2}$.  We
therefore adopt the notation
\begin{equation}
\label{eq:nosub}
\Psi \equiv \Psi_{R,2}.
\end{equation}

Our primary interest is in the marginal case $\sigma=2$, though other
values are also discussed briefly.  The importance of $\sigma=2$ can
be gauged if the Hamiltonian is considered as a quantum-mechanical
system in imaginary time $t=i z$.  The kinetic energy and the $r^{-2}$
interaction (like the angular momentum) scale in the same way for all
dimensions.  The $r^{-2}$ potential is the centrifugal barrier in
quantum mechanics, while it occurs in surface problems as an effective
interaction between steps induced by the elastic strains on the
surface\cite{kohn}.

\subsection{Known exponents for the short range case}
\label{sec:Known-expon-short}
For the noninteracting problem, $v_0=g=0$, the exponents follow from
the Gaussian behavior of the chains. The partition function for a
Gaussian chain of length $N$ is given by
\begin{equation}
  \label{eq:33}
  Z({\bf R} N|{\bf 0} 0) = (2\pi N)^{-d/2} \exp(-R^2/2N),
\end{equation}
from which one obtains
\begin{equation}
  \label{eq:3}
  \psi_{{\rm S},p}=0,\quad \psi_{{\rm R},p}=\frac{pd}{2}, \quad {\rm and}\quad
  \Psi_{{\rm R},p}=\frac{(p-1)d}{2}.
\end{equation}
It is also known that for purely repulsive walkers (vicious walkers) in $d=1$
(i.e., with $g=0$)\cite{fisher84,huse84},
\begin{equation}
  \label{eq:4}
  \psi_{{\rm S},p}=\frac{p(p-1)}{4}, \  \psi_{{\rm
      R},p}=\frac{p^2}{2}, \  {\rm 
    and}\quad \Psi_{{\rm R},p}=\frac{p^2-1}{2}.   
\end{equation}
We earlier generalized the above results to $d=2-\epsilon$ for the
short range case (i.e. $g=0$), by using RG and
obtained\cite{smjpa,smpre}
\begin{mathletters}
\begin{equation}
  \label{eq:5}
    \psi_{{\rm S},p}=\eta_p, \ \psi_{{\rm R},p}=\frac{pd}{2} +2\eta_p,
    \ {\rm
    and}\  \Psi_{{\rm R},p}=\frac{(p-1)d}{2} + 2 \eta_p,\label{miss1}   
\end{equation}
where, with  $\epsilon=2-d$,
\begin{equation}
  \label{eq:31}
  2\eta_p={p\choose 2} \epsilon + 3 {p\choose 3} \ln (3/4) \epsilon^2 +... .
\end{equation}
\end{mathletters}
For $d=2$, our RG yields 
${\cal Z}_{{\rm R},p} \sim N^{-(p-1)} (\ln N)^{-p(p-1)}$, 
which has since  been obtained by the exact
lattice calculations of Essam and Guttmann\cite{gutt} for $p=2$.  This
lends further support to the validity of the renormalization-group
approach.

That only one anomalous exponent is needed was shown explicitly in
Refs. \cite{smjpa,smpre}.  This could be understood from the geometry that a 
reunion  partition function can be thought of as two survival partition 
functions connected together.

In this paper, we calculate the anomalous part $\eta_p$ for the case
with long range interaction at the one loop level. Since this one-loop RG
generates exact results\cite{kolo}, we obtain exact
exponents with nonzero $g$.

\section{Renormalization Group}
\label{sec:Renorm-Group}
\subsection{Survival partition function}
\label{sec:Surv-part-funct}
To evaluate the survival exponent, we use a momentum-shell
RG approach as used in Ref. \cite{kolo}.  The
renormalization process requires integrating out small-scale
fluctuations, so that the effect of interactions of the two chains
within a small distance is taken into account by redefining the
parameters of the Hamiltonian.  In this problem, in addition to the
parameters of the Hamiltonian, we need to consider the survival
partition function also.

In Fourier space, the renormalization of the interaction can be carried 
out using the effective interaction at the one-loop level
\begin{equation}
  \label{eq:8}
  V_{\rm eff}(k)=V(k) - \int^{\Lambda} \frac{d^dq}{(2\pi)^d} \ 
  \frac{V(\mid {\bf k-q}\mid)V(q)}{q^2}.
\end{equation}
The Fourier transform defined generically as
\begin{equation}
  \label{eq:34}
  F({\bf k}) = \int d^d r \ e^{i\bf k\cdot r} G({\bf r}) 
\end{equation}
requires an appropriate analytic
continuation of the integral for singular potentials. 
 In particular, the Fourier
transform of $ r^{-\sigma}$  has a singular part  
\begin{mathletters}
\begin{equation}
  \label{eq:43}
A  \ \frac{k^{-d+\sigma}}{d-\sigma} \quad {\rm in \  general,}
\end{equation}
where
\begin{equation}
  A=2^{d-\sigma+1} \pi^{d/2} \frac{\Gamma((d-\sigma+2)/2)}{\Gamma(\sigma/2)},
\end{equation}
and, for $\sigma=2$, this singular part takes a simpler form,
\begin{equation}
 \frac{1}{K_d} \frac{ k^{2-d}}{d-2},
\end{equation}
\end{mathletters}
where $K_d= S_d/(2\pi)^d$,  $S_d = 2\pi^{d/2}/\Gamma(d/2)$ being the
surface of a $d$-dimensional unit sphere.  A natural choice for the
potential is therefore
\begin{equation}
  \label{eq:42}
  V({\bf k}) = v_0 + \frac{A}{d-\sigma}\  g \ k^{\sigma-d}.
\end{equation}
Let us also introduce at this point  the dimensionless parameters 
\begin{equation}
  \label{eq:35}
\tilde u = v L^{\epsilon}, \ 
\tilde g = K_d {A} g L^{2-\sigma}, \ {\rm and} \ u =
K_d \tilde u + \frac{\tilde g}{d-\sigma}
\end{equation}
with $v=v_0$ as the bare value of the short range coupling
constant. Note that for $\sigma=2$, $g=\tilde g$.

By implementing a thin-shell integration over 
$\Lambda-d\Lambda<k<\Lambda$ and then rescaling the 
cutoff back to $\Lambda \ (=1)$, the
coupling constants' renormalization can be written in the form of 
recursion relations  as 
\begin{mathletters}
\begin{eqnarray}
  \label{eq:9}
  L \frac{du}{dL}&=& \epsilon u - u^2 + \tilde g,\\
  L \frac{d \tilde g}{dL} &=& (2-\sigma) \tilde g.\label{eq:9b}
\end{eqnarray}
\end{mathletters}
The thin-shell integration is analytic and therefore does not
renormalize a singular $g$-type long range potential in the
Hamiltonian.  The flow equation of Eq. (\ref{eq:9b}) for $g$ then
follows from dimensional analysis. The exactness of Eq. (\ref{eq:9}) for
$g=0$ and for $g\ne 0$ is discussed in Refs. \cite{smbjj,kolo}. This
marginality of $g$ is a reflection of the importance of an $r^{-2}$
potential in the Hamiltonian, as is well known in quantum mechanics.
In the surface context, this marginality implies that the strength of
interaction between two steps is invariant under scale transformation.
{\it Henceforth we consider $\sigma=2$, and use $g$ instead of $\tilde
g$}.

 The flow equation for $u$ gives rise to two fixed points
\begin{eqnarray}
u^*_{s,u}=\frac{1}{2}[\epsilon\pm\sqrt{\epsilon^2+4 g}],\label{fixed}
\end{eqnarray}
 where the subscripts
$s$ and $u$ represent stable and unstable fixed points, respectively.

\begin{figure}[htbp]
  \begin{center}
    \narrowtext
   \psfig{file=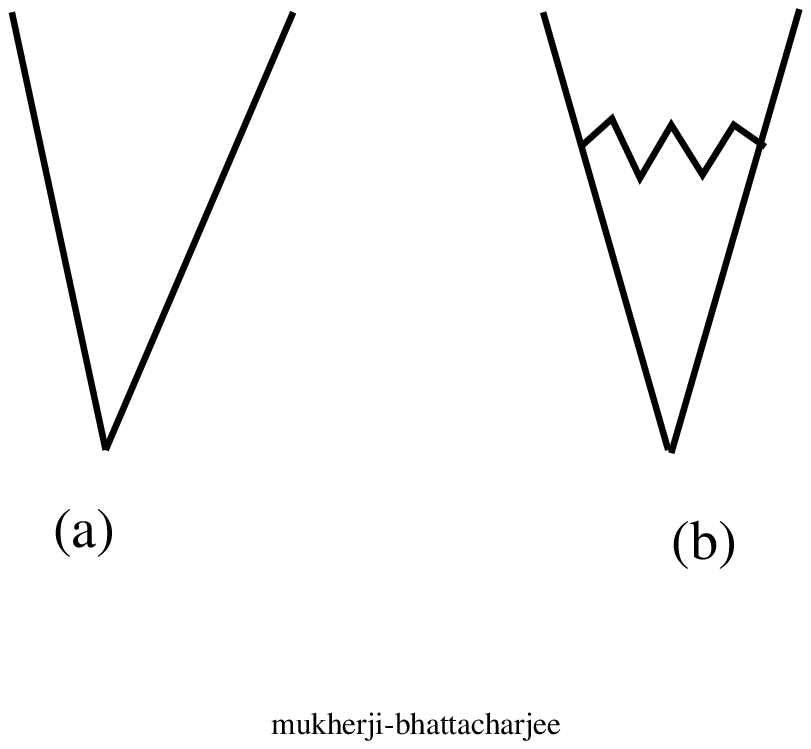,width=3in,angle=270}
    \caption{Zeroth-  and first-order diagrams for survival.  The solid
      lines indicate the polymers and the wavy line interaction.} 
    \label{fig:1}
  \end{center}
\end{figure}

For the survival partition function, the one-loop contribution is
shown in Fig. \ref{fig:1}. The loop contribution is similar to the one-loop
case for the interaction, though here it is $O(u)$. The flow equation
for $Z_{\rm S}$ is (see Appendix A for details)
\begin{equation}
  \label{eq:10}
  L \frac{dZ_{\rm S}}{dL}= - u Z_{\rm S}
\end{equation}
so that at the  stable fixed point of Eq.(\ref{eq:9})
\begin{equation}
  \label{eq:30}
 \eta_2=u_s^*/2. 
\end{equation}

This result can be generalized to any number of chains for which there
will be a combinatorial factor $p\choose 2$ from the pairwise
interaction to yield
\begin{equation}
  \label{eq:11}
  \eta_p=\frac{1}{2} {p\choose 2} u_s^*.
\end{equation}
This exact relation for $p=2$ is a consequence of the connection
between the coupling constant $u$ (vertex function) and the survival
partition function.  The interaction $u$ can be thought of as two
$p=2$ survival partition functions joined at the node so that
renormalization of $u$ is a product of the renormalizations of two
survival partition functions.

\vbox{
\begin{figure}[htbp]
  \begin{center}
    \narrowtext
   \psfig{file=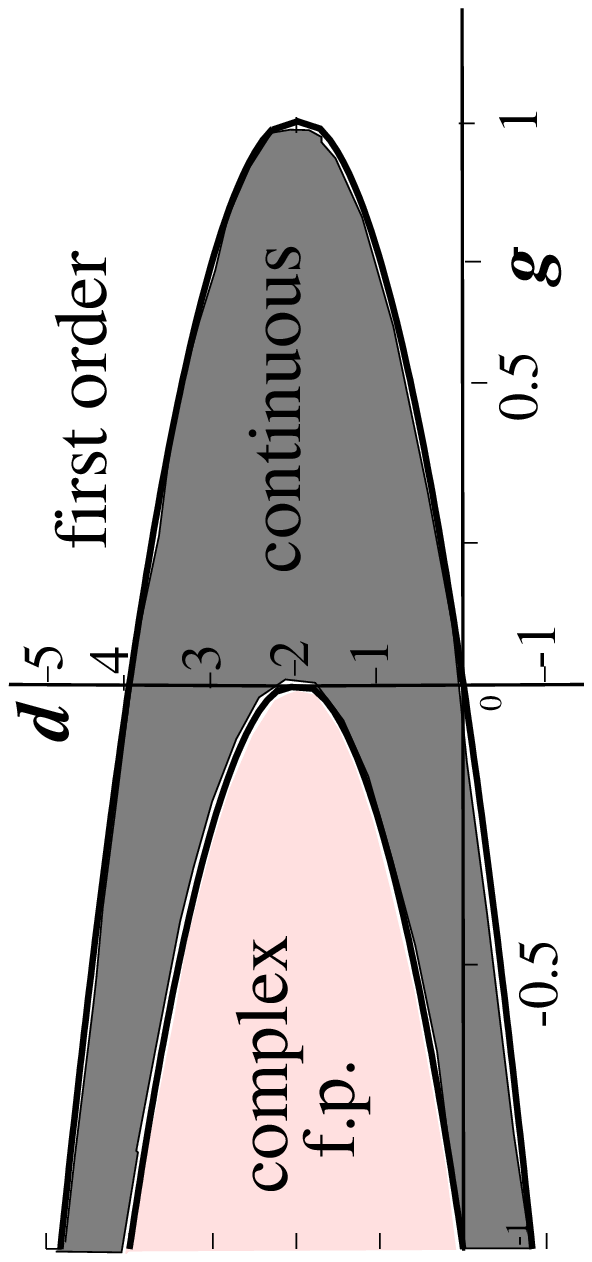,width=3in,angle=270}
    \caption{The locus of points in the $d$ vs $g$ plane for reunion exponent
     $\Psi_{R,2} =1$ and $ \Psi_{R,2} =2$. The shaded region is the region 
      for continuous transition. The transition is first-order outside it.
      The lightly shaded region has complex-conjugate fixed points
      that are of importance in the context of 
      non-hermitian Hamiltonians.}
    \label{fig:dvsg}
  \end{center}
\end{figure}
}

The reunion-anywhere exponent for $d=1$ is  given by 
\begin{equation}
  \label{eq:12}
  \Psi_{{\rm R},p} = \frac{p-1}{2} + {p\choose 2} u_s^*.
\end{equation}
A special case of this is $\Psi_{R,2} = \frac{1}{2} + u_s^*$.  For
only short range interactions  (i.e., $g=0$), $u_s^*=1$ and one gets
back the vicious walker exponent of Huse and
Fisher [$\Psi_{R,p}=(p^2-1)/2$] \cite{fisher84,huse84}.  One notes that
the anomalous part is just the combinatorial factor. The general
results for all three reunion and survival exponents are obtained
by substituting Eq. (\ref{eq:11}) for $\eta_p$ in Eq. (\ref{miss1}).  We
quote the exact result for $d=2-\epsilon$ and $p=2$,
\begin{equation}
  \label{eq:36}
  \Psi\equiv \Psi_{R,2} = \frac{d}{2} + \frac{\epsilon + 
\sqrt{\epsilon^2+4 g}}{2} 
  = 1 + \frac{\sqrt{\epsilon^2+4 g}}{2}.
\end{equation}
This is used in the following sections.

A few things are to be noted here.  First, the fixed points become complex
for ${ g} < -\epsilon^2/4$, in particular for ${ g} <-1/4$ for $d=1$.
For this borderline case $d=2\pm 2 \ \sqrt{- g}$, the reunion
exponent is $ \Psi_{R,2} =1$.  The locus of the points in the $d$ vs
$g$ plane for $\Psi_{R,2} =1$ and $ \Psi_{R,2} =2$ are shown in Fig.
\ref{fig:dvsg}. As we show in Sec. IV, these lines are very special.

\section{Binding-Unbinding transition of polymers}
\label{sec:Bind-Unbind-transiti}
Let us consider two directed polymers, the $p=2$ case of Eq.
(\ref{eq:1}).  The fixed-point diagram of Fig. \ref{fig:3} shows that
at $d=1$, for a given $ g >-1/4$, there will be a binding-unbinding
transition as $v$ is changed (see Fig. 1 of Ref. \cite{smprl99}).
The transition is defined by the unstable fixed point while the stable
fixed point describes the ``high-temperature'' phase.  Several
features of this transition have been discussed in Ref. \cite{kolo} in
the renormalization group framework and in Ref. \cite{lipowsky0}.  We
show here that certain unresolved issues (e.g., order of transition) in
the RG framework can be sorted out by using the reunion exponent,
thereby gaining a complete picture of the problem.  The problem can be
anticipated by noting that the exponent $\nu$ associated with the
diverging length scale, as the transition is reached, approaches the
limit $\nu=1/2$ as $ g\rightarrow 3/4$.  Care should, therefore, be
exercised in drawing conclusions from the flow diagram. We discuss now
how the behavior at the stable fixed points determines the phase
transition at the unstable fixed point.

For generality, we develop the procedure in a general way and then
derive the results for the $1/r^2$ interaction, especially for $p=2$
(two-chain case).  The internal consistency is also shown by a
finite-size scaling argument.  That the exact results are recovered
corroborates the utility of the RG approach, coupled with the reunion
behavior, for other interactions as well.

\subsection{General approach}
\label{sec:General-approach}
\begin{figure}[htbp]
  \begin{center} 
    \narrowtext
    \psfig{file=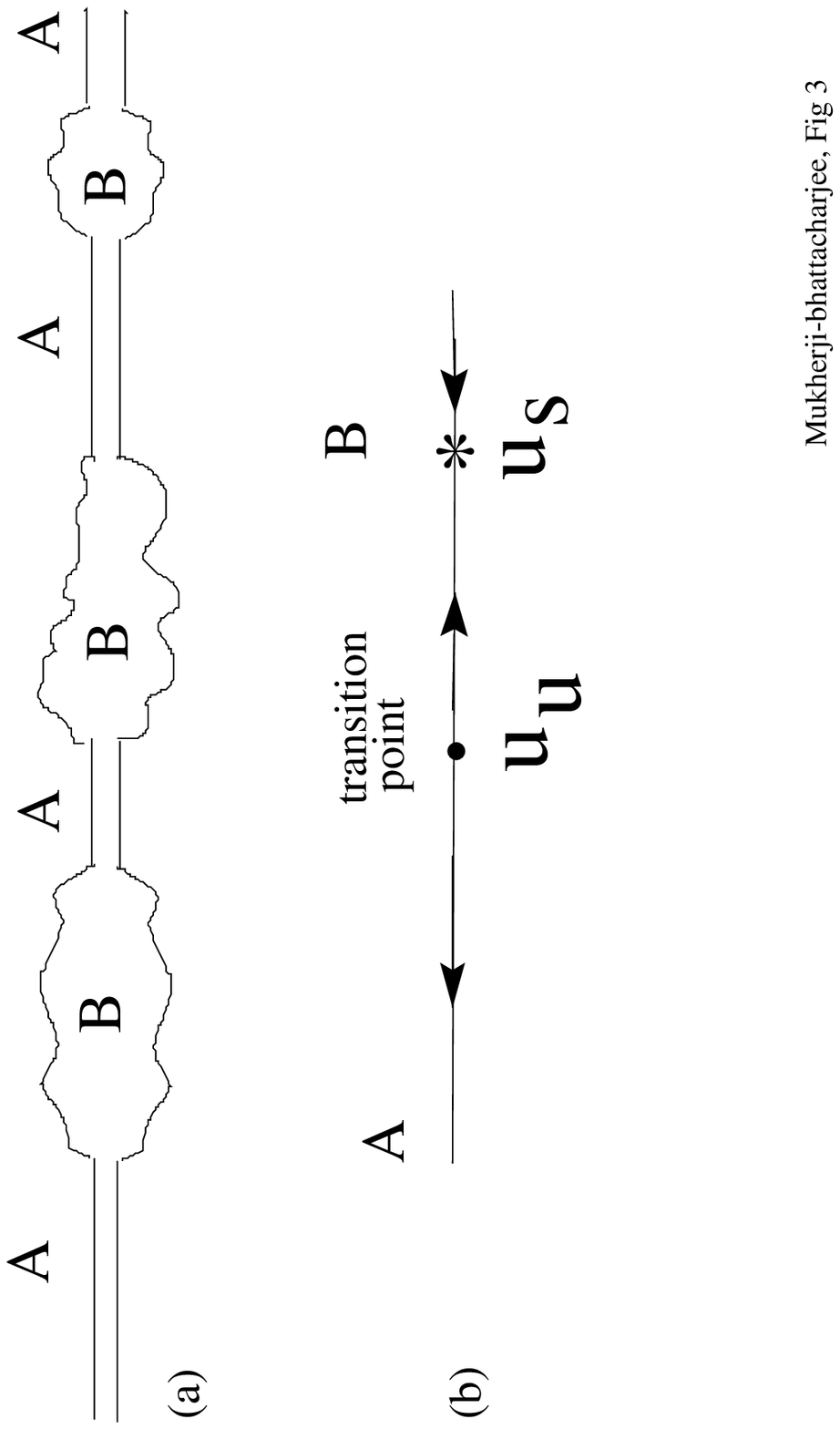,width=3in,angle=270} 
     \caption{(a) A general configuration of the two polymers, 
       consisting of bound (A)
    and unbound (B) regions. (b) A generic fixed-point diagram for $u$
    from the RG flow.  See also Fig. 1 of Ref. [7]. }
\end{center}
\label{fig:3}
\end{figure}

In order to study the nature of the transition, we adopt the procedure
of Ref. \cite{fisher84}.  The partition function is obtained by
summing over all configurations of the two chains. Our main interest
is in the phase-transition behavior and therefore only long-distance
behaviors will be considered.  The configurations can be characterized
by the alternate sequence of domains of bound (A) and unbound (bubble,
B) regions as shown Fig. \ref{fig:3}.  The bound regions
correspond to the left side of the unstable fixed point of the flow
diagram (the coupling constant flows to $-\infty$), while the bubbles
correspond to the high-temperature phase characterized by the stable
fixed point.

Let us denote the partition function for a segment of length $N$ of
type X by $Q_N^{\rm X}$.  We assign a weight $v$ to each A-B and B-A
junction, and for simplicity we consider cases in which  the chains start and
end in A-type domains.  This extra factor $v$ is to take care of the
deviation from the strict A- or B-type behavior at the junctions. The
total partition function can be written as
\begin{equation}
  \label{eq:13}
  Q_N^{\rm T}=Q_N^{\rm A} + \int_0^N \!\! dz_2 \ \int_0^{z_2} \!\!
  dz_1 Q_{N-z_2}^{\rm A} 
  v\ Q_{z_2-z_1}^{\rm B} v\  Q_{z_1}^{\rm A} + ... \ .
\end{equation}
To exploit the convolution nature of the terms of the series, we go
over to the grand-canonical ensemble with respect to the chainlength.
This is equivalent to taking the Laplace transform of the partition
functions with respect to $N$.  Defining the Laplace transform as
\begin{equation}
  \label{eq:14}
  G^{\rm X} (s) = \int_0^{\infty} e^{-sN} Q_N^{\rm X} \ dN,
\end{equation}
where X is A, B or T, the above series can be written in the form of a
geometric series and can be summed to obtain
\begin{equation}
  \label{eq:15}
  G^{\rm T}(s)= \frac{G^{\rm A} (s)}{1-v^2 G^{\rm A}(s)G^{\rm B}(s)}.
\end{equation}
The existence of the thermodynamic limit ensures that $G^{\rm X}(s)$
is analytic for $s >f_X$, where $f_X=\lim_{N\rightarrow \infty} N^{-1}
\ln Q_N^{\rm X}$, and $G^{\rm X}(s) \rightarrow 0$ as $s \rightarrow
\infty$.  In other words, the largest real singularity of $G^{\rm
X}(s)$ in the complex-$s$ plane determines the corresponding limiting
free energy per unit length\cite{comment1}.  For $G^{\rm T}(s)$, this
could either be the singularity of $G^A$ or $G^B$ or may come from the
vanishing of the denominator of Eq. (\ref{eq:15}), i.e. the solution of
\begin{equation}
  \label{eq:17}
  v^2 G^{\rm B}(s) = 1/G^{\rm A}(s).
\end{equation}
In case $G^A$ or $G^B$ diverges, then the denominator of Eq.
(\ref{eq:15}) vanishes for a larger value of $s$, and then it is the
root of Eq. (\ref{eq:17}) that is the relevant one.  In such a
situation, the free energy will not have any singularity. It follows
that it is only the nondiverging branch points that could lead to
phase transitions.

For the A-type region, the partition function $Q_N^{\rm A}= e^{-\beta
\varepsilon N}$, where $\beta$ is the inverse temperature. Its Laplace
transform has a simple pole.  Therefore, the singularity of $G^A$ is
not important for our discussion.  Consequently, it is the partition
function of the bubbles (high-temperature phase) that determines the
behavior of the chains\cite{lifson,polsch,fisher65}.  The partition
function of the bubble at the stable fixed point for large $N$ is
given by
\begin{equation}
  \label{eq:18}
  Q_N^B\approx \frac{q e^{-N\sigma_0(T)}}{N^{\Psi}},
\end{equation}
where the reunion-anywhere exponent $\Psi$ at the stable fixed point
appears.  Here and in the following discussion, for convenience, we
have removed the subscripts of the reunion exponent introduced in
Sec. II [see Eq. (\ref{eq:nosub})].  The Laplace transform of this
partition function has a singularity at $s=\sigma_0(T)$, and the
nature of the singularity depends on the value of the reunion
exponent, $\Psi$.  There is a divergence at this singularity if
$\Psi<1$.  A phase transition therefore occurs only for $\Psi \ge 1$.
The details of the graphical solution, for the discrete case, can be
found in Ref. \cite{fisher84}.

For $\Psi>1$, the high temperature phase is described by the root of
Eq. (\ref{eq:17}) and the transition is given by the temperature $T=T_c$
at which the root of Eq. (\ref{eq:17}) coincides with the singularity of
$G^B(s)$.  Defining the deviation of temperature as $t=T_c-T$, the
free energy [from the root of Eq. \ref{eq:17}] can be written as
\begin{mathletters}
\begin{eqnarray}
  \label{eq:20}
 f_{\rm T}&\approx &\sigma_0(T)-  \mid t\mid^{1/(\Psi-1)}\  \hfill {\rm for
    }\quad 1<\Psi<2,\\
&\approx &\sigma_0(T)- \sum_{1}^{m-1} a_j t^j + \mid
t\mid^{(\Psi-m)}\nonumber\\
&& \hfill {\rm for  }\quad m<\Psi<m+1.
\end{eqnarray}
\end{mathletters}
We find a critical
behavior for $1<\Psi<2$ but a first-order transition for $\Psi >2$.
In case of a critical behavior, the specific-heat exponent can be read 
off from the free energy as
\begin{equation}
  \label{eq:21}
  \alpha= 2- (\Psi -1)^{-1}= \frac{2\Psi -3}{\Psi -1}.
\end{equation}

\subsection{d=1}
\label{sec:d=1}
The crucial feature in the approach developed in the preceding
subsection is the alternate sequence of the two types of ``bubbles''
and therefore the results can be used, for example, in the case of
unbinding of $p$ chains with $p$-body interaction\cite{smbjj,smpre}
and other cases.  To take advantage of previous studies, we
concentrate on the two-chain problem.

Let us now consider the case of two directed polymers in
$(1+1)$ dimensions, for which we obtained $\Psi=(1+\sqrt{1+4 g})/2$.
The fixed points (see Fig. \ref{fig:3}) indicate that there is a
binding-unbinding transition at the unstable fixed point for any
given $g$ as $u$ is varied.  For any $u_0 < u_{\rm u}^*$, the
renormalized flow goes to $-\infty$, so that a length scale can be
identified from $u(l^*) \rightarrow-\infty$.  From this, as $u_0
\rightarrow u_{\rm u}^*$, the diverging length scale comes out as
($t\equiv \Delta u= u - u_u^*$)
\begin{equation}
  \label{eq:22}
  \xi_{\perp}\sim \mid t\mid^{-\nu_{\perp}} \quad\quad {\rm
    with}\quad \nu_{\perp} = \frac{1}{u_s^* -u_u^*}.
\end{equation}
For $d=1$, we have 
\begin{equation}
\nu_{\perp} = \frac{1}{\sqrt{1+4 g}}.  
\end{equation} 
This shows that $\nu_{\perp}=1/2$ at $ g=3/4$, a signature of
something special happening there.  The reunion exponent, coupled with
the analysis of the preceding subsection, tells us that there cannot
be a phase transition for $ g<-1/4$, a fact corroborated by the
fixed-point diagram (see Fig. 1 of Ref. \cite{smprl99}).  A critical
behavior is expected for $3/4 \ge g\ge -1/4$ and a first-order
transition for $ g>3/4$.

For $-1/4\le g\le 3/4$, which includes the pure short range $ g=0$
case, the average fraction of the length in the bound state is given
by
\begin{equation}
  \label{eq:23}
  \Theta(t)\sim\frac{\partial f}{\partial t} \sim \mid
  t\mid^{\beta}.
 \end{equation}
 This defines the order-parameter exponent $\beta$. This exponent and 
 the specific-heat exponent are given by
\begin{equation}
    \label{eq:39}
   \beta=\frac{2-\Psi}{\Psi-1}=\frac{2-\sqrt{1+4 g}}
{\sqrt{1+4 g}},   
\ {\rm and}\ \alpha = 2\frac{\sqrt{1+4 g}  -1}{\sqrt{1+4 g}}.
\end{equation}
Note that $\alpha=1$ at $ g=3/4$, a requirement for a
first-order transition.

The bubble lengths have fluctuations and this gives the measure of the 
diverging length scale parallel to the chain as
\begin{equation}
  \label{eq:24}
    \xi_{\parallel}\sim \mid \!t\!\mid^{-\nu_{\parallel}} \quad\quad {\rm
    with}\quad \nu_{\parallel} =2 \nu_{\perp}. 
\end{equation}
Note that the hyperscaling relation
$d\nu_{\parallel}=2-\alpha$ is obeyed by these nonuniversal exponents 
with $d=1$.
This is because the free-energy density is the free energy per unit
length of the polymers and not the unit $(d+1)$-dimensional area.

If $ g>3/4$, $\Theta(t) \rightarrow$ const. as $t\rightarrow 0-$.
This indicates a first-order transition. The longitudinal length-scale
exponent is $\nu_{\parallel}=1$, i.e., it sticks to its value at
$g=3/4$.  However, the free energy as given by Eq. (\ref{eq:20}) shows a
weak singularity that will be reflected in the divergence of an
appropriate higher derivative.  One therefore finds a rather unusual
first-order transition with weak singularities and diverging length
scales\cite{lipowsky0}.  Such scales can be determined from the higher
cumulants of the length fluctuations of the bubbles. We discuss these
issues in the next section in the context of the equivalent quantum
problem.
 
$\Psi$ increases with $g$ and every-time $\Psi$ crosses an integer,
the diverging derivative shifts by 1.  These special values are
$g=(m-1)^2- 1/4$.

At $ g=-1/4$,  two fixed points $u_s^*$ and $u_u^*$ merge at a 
common value $u^*=1/2$. The bound phase corresponds to the
$u_0<1/2$ region and the transverse length scale diverges 
as 
\begin{eqnarray}
\xi_\perp \sim \exp[1/(1/2-u_0)].
\end{eqnarray}
However a  Kosterlitz-Thouless-type behavior is observed if $g$ is varied
\cite{kolo}.

\subsubsection{Finite size scaling for the critical point}
\label{sec:Finite-size-scaling}
We have derived the exponents for the critical point from our results
of the stable fixed point or the high-temperature phase.  An independent
verification of the exponents comes from a finite-size scaling
argument used strictly in the critical region characterized by the
unstable fixed point.   The reunion describes the contacts 
of the two chains
and therefore the number of contacts at the unstable 
fixed point would have a 
finite-size scaling behavior ($L$ is a length along the chain) 
\begin{equation}
  \label{eq:7}
 \Theta_L \sim L^{-\Psi^{\{u\}}} \sim t^{\Psi^{\{u\}}/\nu_{\parallel}},
\end{equation}
which identifies
\begin{equation}
  \label{eq:37}
  \beta=\frac{2-\Psi^{\{s\}}}{\Psi^{\{s\}}-1}=
\frac{\Psi^{\{u\}}}{\nu_{\parallel}},
\end{equation}
where we have introduced the superscripts \{s\} and \{u\} to
distinguish the values of the exponent at the stable and unstable
fixed points, respectively. [$\Psi$ of Eq. (\ref{eq:39}) is
$\Psi^{\{s\}}$ here.]  Using the unstable fixed point value in
Eq. (\ref{eq:12}), we do see this equality to be true because
\begin{equation}
  \label{eq:38}
  \Psi^{\{u\}} = 1 - \frac{\sqrt{1+4 g}}{2}.
\end{equation}

\subsection{$d\neq 1$}
\label{sec:d=2-epsilon}
For $1<d<2$, i.e., $d=2-\epsilon$ with positive $\epsilon$, the
results of the preceding subsection can be repeated by using $\Psi$ of
Eq. (\ref{eq:36}).  The reunion exponent requires $ g>-\epsilon/4$ (so
that $\Psi>1$) for a phase transition that is identical to the
condition of the real fixed point of the flow equation. The length-scale
exponent becomes equal to $1/2$ at $ g=d(4-d)/2=(4-\epsilon^2)/2$,
which coincides with the value of $g$ at which $\Psi =4$.  The
exponents satisfy the general scaling relations of Eqs.
(\ref{eq:24}) and (\ref{eq:37}).

The stability of the fixed points flips at $d=2$.  The results for
$d>2$ can be obtained by doing an analytic continuation from $\epsilon
>0$ to $\epsilon <0$.  For example, on the stable branch now $\Psi= 1
+ \sqrt{\epsilon^2 + 4g}/2$.  Figure \ref{fig:dvsg} shows the regions of
critical and first-order transition in the $d$-vs-$g$ plane.

\section{Quantum problem}
 \label{sec:Quantum-problem}
 The Hamiltonian of Eq. (\ref{eq:1}) represents two quantum particles
 either in the path-integral representation in imaginary time or as a
 statistical mechanical problem at $T=0$. In the center-of-mass frame,
 the equivalent Schr\"odinger equation for $\sigma=2$, where $v_0(r)$ is the 
short range part of the interaction, is 
\begin{equation}
   \label{eq:25}
   -\nabla^2\psi({\bf r}) + \left [ v_0({\bf r}) + \frac{g}{r^2}\right ]
   \psi({\bf r}) = E \psi({\bf r}).
\end{equation}
For zero energy, the radial part of the wave function can be written as 
\begin{equation}
   \label{eq:26}
   R(r)= \exp\left [\int \frac{u(r)}{r} dr\right] 
\end{equation}
 with $u(r)$ satisfying
 \begin{equation}
   \label{eq:27}
   r \frac{du}{dr} = (2-d) u - u^2 + g.
\end{equation}
Taking $l=\ln(r/a)$, we can recast the above equation in a form
resembling the RG flow equation.  In other words, the RG flow equation
at long distances determines the radial part of the zero-energy
wave function.

We consider the transition point itself, which corresponds to the
unstable fixed point.  In case $u$ of Eq. (\ref{eq:27}) reaches a fixed
point, the wave function has an algebraic tail as 
$R(r) \sim \mid r\mid^{u^*}$.

The wave function $R(r)$ satisfying the Schr\"odinger equation is
analogous to the partition function of a polymer starting from origin
to the point $r$.  Its behavior is similar to the survival partition
function because in the process the particle may feel (in a
perturbative approach) the potential any number of times.  The scaling
that time and space are related by $[N]=[r]^2$ then gives, via the
scaling of the survival partition function, $R(r) \sim \mid
r\mid^{2\Psi_{S,2}} \sim \mid r\mid^{u^*}$.  This is what we get from
Eq. (\ref{eq:27}).
   
At $d=1$, we then get at $u=u_{\rm u}^*$ [the unstable fixed point of
Eq. (\ref{fixed}) with $\epsilon=1$] 
$R(r) \sim \mid r\mid^{u_{\rm    u}^*}$.  The moments of scalar 
$r^2={\bf r}\cdot{\bf r}$, defined for general $d$ by
\begin{equation}
   \label{eq:28} 
<r^{2p}>=\frac{\int ({\bf r}\cdot{\bf r})^p R(r)^2
   r^{d-1}dr}{\int R(r)^2 r^{d-1} dr},
\end{equation} 
would depend on the long tail of the wave function.  The wave-function
itself is not normalizable for $u_{\rm u}^* >-1/2$, i.e., for $g<3/4$.
This is a signature of an unbound state.  In such a situation, by
tuning the short range part of the potential from below, one can get a
bound state arbitrarily close to zero energy.  As the zero-energy
state is reached, the length scale that measures the boundedness or
localization of the state increases without bound. Obviously, all the
moments are divergent.  This is the quantum picture of the criticality
discussed in the preceding section for polymers.
 
For $g>3/4$, the zero-energy wave function is normalizable.  This
means a bound state with zero energy but with a long tail of the
wave function.  In this region, the bound state can also be made to
approach the zero energy state but there will always be a finite
length scale coming from the finite moments of the wave-function.
This is a first-order transition, but with a difference.  The moments
of $r^2$ are finite for $p<(\sqrt{1+4g})/2 - 1$, i.e., for $p<\Psi
-2$, and any value of $p < \Psi -2$ may be used to define a finite
length scale.  However, the divergence of the higher moments indicates
some remnants of the ``criticality.'' The integer values of the
reunion exponent, which show up in the free energy in Eq.~(\ref{eq:20}),
also make their presence felt here as the special values at which a
new integer moment becomes finite.

We immediately see that all the moments of $r^2$ are diverging for 
${u_{\rm u}^*}>-1/2$, and the first $p$ moments are finite for 
\begin{equation}
\label{moment}
 g > (p+1)^2-1/4.
\end{equation} 
For $p$ violating this criterion, using a finite-size scaling analysis (i.e., 
by cutting off the integral for large $r$ by the length scale), one gets
\begin{equation}
\label{num}
\nu_p=\frac{1}{4p}\ (2p+2 - {\sqrt{1+4g}}).
\end{equation}
That there is a diverging length scale coming from
higher moments gives an {\it a posteriori} justification of the RG
analysis based on fixed points.

The states with $E<0$ always have at least exponential (or faster)
decays of the wave function at large distances.  This rapid decay
ensures finiteness of all moments.  However, at the transition point,
the possibility of diverging moments arises because of the power-law
(``critical'') decay of the wave function.

\subsection{Short range interaction, general $d$}
\label{sec:d>1,-short-range}
We now show that the one-dimensional exact results for the reunion
behavior can be made to bear upon the higher-dimensional problems
also.  In the process, we recover many results obtained earlier
from detailed exact solutions
for each $d$ by using the properties of special 
functions\cite{balian,lipowsky}.

Let us consider a short range central potential $V({\bf r})$.
Defining $\phi(r)=r^{(d-1)/2} R(r)$, where $R(r)$ is the radial part
of the wave function, the Schr\"odinger equation in $d$ dimensions can
be written as\cite{balian}
\begin{equation}
  \label{eq:32}
 - \frac{d^2\phi}{dr^2} + \left (\frac{A_l(d)}{r^2} + V(r)\right )
 \phi(r) = E \phi, 
\end{equation}
where $A_l(d) = (d+2l-3)(d+2l-1)/4$ is the coefficient of the
angular momentum ( or centrifugal) barrier, $l$ being the integer
angular momentum quantum number.  Note that the factor $r^{(d-1)/2}$
makes the integral over $r$ the same as that of the one-dimensional
problem.  The $d$-dimensional problem is then reduced to the one
dimensional problem and the exact results for $d=1$ for various $ g$
as obtained in the previous subsection can be used to get the features
as $d$ and $l$ are changed.  The ground state is obtained from $l=0$
for $d \ge 1$.  For other values of $d$, $l$ is to be chosen such that
$1-2l\le d<3-2l$.  (The choice makes the minimum centrifugal
barrier.)  By using our results, we see that for $A_l(d) < 3/4$, there
is no zero-energy bound state, though the state can be reached
continuously from below.  Substituting the expression for $A_l(d)$, we
find $\Psi=\frac{d}{2} + l$ for $d>2$.  This shows that for $l=0$ and
$1<d<4$, there can be no bound state at zero energy.  As the
short range parameter is tuned, the localization length of the bound
state diverges with an exponent
\begin{equation}
  \label{eq:19}
  \nu_{\perp} = \frac{1}{2} \nu_{\parallel} = \frac{1}{\mid d-2\mid}
  \quad {\rm for }\ \ 1\le d< 4,
\end{equation}
with $\Psi=2$ at $d=4$.  For $d<1$, the ground state comes from
$l=1$, so that
\begin{equation}
  \label{eq:40}
    \nu_{\perp} = \frac{1}{\mid d\mid}
  \quad {\rm for }\ \ -1\le d< 1.
\end{equation}
Similarly, $\nu_{\perp}=1/\mid d+2\mid$ for $-3\le d <-1$ when $l=2$.
In general,
\begin{equation}
  \label{eq:41}
  \nu_{\perp}  = \frac{1}{\mid d+2l-2\mid}
  \quad {\rm for }\ \ 1-2l\le d< 3-2l.
\end{equation}
For all $d<4$, $\nu_{\parallel}/\nu_{\perp} = 2$.  For $d >4$, we find  
from Eq. (\ref{moment}) that the  $p$th moment of $r$ will be finite
\cite{lipowsky} if $ p<d-4$, and for $p>d-4$ one gets 
\begin{equation}
\nu_p=\frac{1}{2} - \frac{d-4}{2p}.
\end{equation}

\section{Conclusion}
In conclusion, we have investigated the reunion of random walkers
having both short range and long range interactions. By using a
momentum-shell renormalization-group technique, the reunion and
survival exponents have been calculated.  The exponent $\Psi\equiv
\Psi_{R,2}$ for polymers has been evaluated at the unbound phase and
at the binding-unbinding transition point represented by the stable
and the unstable fixed points, respectively, in the coupling constant
space. The value of this exponent in the unbound phase is crucial in
determining the nature of the binding-unbinding
transition\cite{lipowsky0,kolo,smprl99}. This transition is critical
for $1<\Psi<2 $ and first order with higher moments diverging for
$\Psi>2$. Since $\Psi$ is explicitly dependent on the strength of the
long range interaction $g$, the order of the phase transition depends
on this parameter.  See Fig. \ref{fig:dvsg}.  For example, at $d=1$,
the dimension more relevant in the context of the experimental
observation of vicinal surfaces\cite{smprl99}, one finds a first-order
transition for $g>3/4$ and criticality for $-1/4<g<3/4$. In the
quantum-mechanical picture, the different nature of the phase
transition is reflected in the approach of the bound state to the
zero-energy state as the short range part of the potential is
tuned. For 
$g<3/4$, one can get the bound state arbitrarily close to the
zero-energy state with a diverging length scale as the gap vanishes,
whereas for the first-order case ($g>3/4$), the length scale remains
finite. In the latter case, it is possible to define diverging length
scales from higher moments [see Eq. (\ref{num})].  This one-dimensional
case can further be extended to the case of a quantum particle
subjected to a short range potential in dimensions $d\neq 1$ with the
centrifugal barrier playing the role of the long range potential.
The fact that all the details and the nuances of the $r^{-2}$ interaction
problem known from exact
solutions\cite{suth,lipowsky0,balian,lipowsky} could be recovered in a
unified manner via the reunion behavior lends credence to the general
approach developed in this paper. Our method can be used for other
problems and interactions as well.

\appendix
\section{Renormalization of the survival partition function}
In the noninteracting case, the survival partition function $Z_s$ for
Gaussian walkers is unity. The anomalous exponent of $N$ in the
survival partition function appears due to the interaction among the
chains. The nontrivial contribution of this interaction is apparent
from the one-loop $O(u)$ term, which is shown in Fig \ref{fig:1}.  The
contribution of this diagram is
\begin{eqnarray}
\int_0^N \int d{\bf r_1}d{\bf r_2}Z(z,{\bf r_1}\mid 0,{\bf 0})
Z(z,{\bf r_2}\mid
0,{\bf 0})\times \nonumber\\
(v_0 \delta({\bf r}_1-{\bf r}_2)+\frac{g}{\mid r_1-r_2\mid^\sigma})
\end{eqnarray}
In the Fourier space, this term appears as
\begin{eqnarray}
\int_0^N \int d{\bf p}\ Z(z,{\bf p})Z(z,-{\bf p})[v_0+v_{\rm{lr}}(p)],
\end{eqnarray} 
where $v_{\rm{lr}}(p)=\frac{A g p^{\sigma-d}}{d-\sigma}$. By using
$Z(z,{\bf p})=\exp[-p^2 z/2]$, we perform integration only over an
outer shell of radii $\Lambda$ and $\Lambda(1-\delta)$, where $\delta$
is a very small parameter. This is done essentially to integrate out
the small-scale fluctuations. In the large-$N$ limit, the contribution
of the above term after the momentum-shell integration is
\begin{eqnarray}
\Lambda^{-\epsilon}\delta K_d[v_0+v_{\rm{lr}}(\Lambda)]
\end{eqnarray}
By taking this term into account, one can express $Z_s$ viewed at a larger
length scale $(\Lambda-\delta\Lambda)^{-1}$ in terms of $Z_s$ viewed at a 
smaller resolution as
\begin{eqnarray}
Z_s(\Lambda-\delta\Lambda)=Z_s(\Lambda)-Z_s(\Lambda)\ K_d\Lambda^{-\epsilon}
[v_0+v_{\rm{lr}}(\Lambda)],
\end{eqnarray}
where only the cutoff dependence is shown explicitly.  This equation
can further be transformed into a differential equation form
\begin{eqnarray}
\frac{d\ln Z_s}{d\Lambda}=-u,
\end{eqnarray}
where $L=\Lambda^{-1}$. By using the fixed-point value for $u$, one
obtains the exponent given in Eq. (\ref{eq:30}).

\end{multicols}

\begin{thebibliography}{99}
\bibitem{comment1}  This is analogous to the condition determining the 
  radius of convergence of the generating function for the discrete
  case of Ref. \cite{fisher84}.  Note that this corresponds to the 
  root closest to origin in the formulation of Ref. \cite{fisher84}.
\end{thebibliography}
\end{document}